%% file: Proceedings-SusHi-EPS-HEP2023.tex
\documentclass[11pt, a4paper, toc=bibliography,DIV=11,numbers=noenddot]{article} 
\pdfoutput=1

\usepackage{pos}

\makeatletter
\DeclareOldFontCommand{\rm}{\normalfont\rmfamily}{\mathrm}
\DeclareOldFontCommand{\sf}{\normalfont\sffamily}{\mathsf}
\DeclareOldFontCommand{\tt}{\normalfont\ttfamily}{\mathtt}
\DeclareOldFontCommand{\bf}{\normalfont\bfseries}{\mathbf}
\DeclareOldFontCommand{\it}{\normalfont\itshape}{\mathit}
\DeclareOldFontCommand{\sl}{\normalfont\slshape}{\@nomath\sl}
\DeclareOldFontCommand{\sc}{\normalfont\scshape}{\@nomath\sc}
\makeatother
\usepackage{amsmath}
\usepackage{scalefnt}
\usepackage{enumitem}
\usepackage[absolute]{textpos}
\usepackage{xcolor}
\usepackage[nice]{nicefrac}
\usepackage{tabularx,booktabs}
\usepackage{xspace}
\usepackage{epstopdf}
\usepackage{listings}
\usepackage{algorithm}
\usepackage{pifont}
\usepackage{algorithmicx}
\usepackage{algpseudocode}
\usepackage[final]{showkeys}
\usepackage{filemod}
\usepackage{bold-extra}
\usepackage[font=small,labelfont={sf,bf},format=plain,margin=0.05\textwidth]{caption}
\usepackage{subcaption}
\usepackage[parfill]{parskip}

\definecolor{gray}{RGB}{125,125,125}
\definecolor{lightgray}{RGB}{190,190,190}
\definecolor{LightRed}{rgb}{1, 0.25, 0.25}
\definecolor{Magenta}{RGB}{255,0,255}
\definecolor{Yellow}{RGB}{226,254,107}
\definecolor{khaki}{rgb}{0.76, 0.69, 0.57}
\definecolor{violet}{rgb}{0.73, 0.2, 0.52}
\definecolor{lilac}{rgb}{0.53, 0.38, 0.56}
\definecolor{gold}{rgb}{0.83, 0.69, 0.22}
\definecolor{orange}{rgb}{1.0, 0.74, 0.53}
\definecolor{mint}{rgb}{0.67, 0.94, 0.82}
\definecolor{maroon}{rgb}{0.69, 0.19, 0.38}
\definecolor{pink}{rgb}{1.0, 0.57, 0.64}
\definecolor{White}{RGB}{255,255,255}
\definecolor{Black}{RGB}{0,0,0}

\usepackage{setspace}

  \usepackage{graphicx}

\newcounter{notecount}

\newcommand{\citere}[1]{Ref.\,\cite{#1}}
\newcommand{\citeres}[1]{Refs.\,\cite{#1}}

\newcommand{\eqn}[1]{Eq.\,(\ref{#1})}

\newcommand{\fig}[1]{Fig.\,\ref{#1}}

\newcommand{\abbrev}[1]{{\scalefont{1}#1}}

\newcommand{\acro}[3]{%
  \newcommand{#1}{#3 (\abbrev{#2})\renewcommand#1{\abbrev{#2}}%
  }%
}

\newcommand{\acroplus}[4]{%
  \newcommand{#1}{#4 (\abbrev{#3})\renewcommand#1{\abbrev{#3}}\renewcommand{#2}{\abbrev{#3}s}%
  }%
  \newcommand{#2}{#4s (\abbrev{#3})\renewcommand#1{\abbrev{#3}}\renewcommand#2{\abbrev{#3}s}%
  }%
}

\newcommand{\acroplu}[5]{%
  \newcommand{#1}{#4 (\abbrev{#3})\renewcommand#1{\abbrev{#3}}\renewcommand{#2}{\abbrev{#3}s}%
  }%
  \newcommand{#2}{#5 (\abbrev{#3})\renewcommand#1{\abbrev{#3}}\renewcommand#2{\abbrev{#3}s}%
  }%
}

\acro{\qcd}{QCD}{Quantum Chromodynamics}
\acroplus{\pdf}{\pdfs}{PDF}{parton density function}
\acroplus{\pl}{\pls}{PL}{Polylogarithm}
\acroplus{\mpl}{\mpls}{MPL}{Multiple Polylogarithm}
\acroplus{\gpl}{\gpls}{GPL}{Generalised Polylogarithm}
\acroplus{\hpl}{\hpls}{HPL}{Harmonic Polylogarithm}
\acro{\sm}{SM}{Standard Model}
\acro{\lhc}{LHC}{Large Hadron Collider}
\acroplus{\bsm}{\bsms}{BSM}{beyond the \sm\ model}
\acroplu{\ff}{\ffs}{FF}{FireFly}{FireFlies}
\acro{\lo}{LO}{leading order}
\acro{\nlo}{NLO}{next-to-leading order}
\acro{\nnlo}{NNLO}{next-to-next-to-leading order}
\acro{\ntlo}{N$^3$LO}{next-to-next-to-next-to-leading order}
\acroplu{\cme}{\cmes}{CME}{centre of mass energy}{centre of mass energies}
\acro{\htl}{HTL}{heavy top limit}
\acro{\thdm}{2HDM}{Two-Higgs-doublet model}
\acro{\mssm}{MSSM}{Minimal Supersymmetric \sm}
\acro{\ew}{EW}{electro-weak}
\acro{\fivefs}{5FS}{five-flavour scheme}
\acro{\fourfs}{4FS}{four-flavour scheme}
\acro{\dy}{DY}{Drell-Yan}
\acro{\slha}{SLHA}{SUSY Les Houches Accord}
\acro{\moca}{MC}{Monte-Carlo}

\newcommand{\sushi}{\texttt{SusHi}}
\newcommand{\sushit}{\texttt{SusHi 2.0}}
\newcommand{\vh}{\texttt{vh@nnlo}}

\newcommand{\feynhiggs}{\texttt{FeynHiggs}}
\newcommand{\thdmc}{\texttt{2HDMC}}
\newcommand{\spheno}{\texttt{SPheno}}
\newcommand{\softsusy}{\texttt{SOFTSUSY}}
\newcommand{\ggh}{\texttt{ggh@nnlo}}
\newcommand{\bbh}{\texttt{bbh@nnlo}}
\newcommand{\zwprod}{\texttt{zwprod}}
\newcommand{\lhapdf}{LHAPDF}
\newcommand{\cuba}{\texttt{CUBA}}
\newcommand{\vegas}{\texttt{VEGAS}}
\newcommand{\lt}{\texttt{LoopTools}}
\newcommand{\fs}{\texttt{FlexibleSUSY}}

\newcolumntype{C}[1]{>{\centering\arraybackslash}m{#1}}
\newcolumntype{R}[1]{>{\raggedleft\arraybackslash}m{#1}}
\newcolumntype{L}[1]{>{\raggedright\arraybackslash}m{#1}}

\def\cpc#1#2#3#4{\href{https://dx.doi.org/#4}{{\it Comp.~Phys.~Commun.~}\jref{\bf #1}{#2}{#3}}}
\def\epjc#1#2#3#4{\href{https://dx.doi.org/#4}{{\it Eur.~Phys.~J.~}\jref{\bf C #1}{#2}{#3}}}

\def\jcp#1#2#3#4{\href{https://dx.doi.org/#4}{{\it J.~Comp.~Phys.~}\jref{\bf #1}{#2}{#3}}}

\def\jhep#1#2#3#4{\href{https://dx.doi.org/#4}{{\it JHEP~}\jref{\bf #1}{#2}{#3}}}
\def\jpcs#1#2#3#4{\href{https://dx.doi.org/#4}{{\it J.~Phys.~Conf.~Ser.~}\jref{\bf #1}{#2}{#3}}}

\def\mpl#1#2#3#4{\href{https://dx.doi.org/#4}{{\it Mod.~Phys.~Lett.~}\jref{\bf A #1}{#2}{#3}}}

\def\nca#1#2#3#4{\ifthenelse{\equal{#4}{}}{{\it Nuovo~Cim.~}\jref{\bf #1A}{#2}{#3}}{\href{https://dx.doi.org/#4}{{\it Nuovo~Cim.~}\jref{\bf #1A}{#2}{#3}}}}
\def\npb#1#2#3#4{\href{https://dx.doi.org/#4}{{\it Nucl.~Phys.~}\jref{\bf B #1}{#2}{#3}}}

\def\plb#1#2#3#4{\href{https://dx.doi.org/#4}{{\it Phys.~Lett.~}\jref{\bf B #1}{#2}{#3}}}

\def\prd#1#2#3#4{\href{https://dx.doi.org/#4}{{\it Phys.~Rev.~}\jref{\bf D #1}{#2}{#3}}}

\def\prl#1#2#3#4{\href{https://dx.doi.org/#4}{{\it Phys.~Rev.~Lett.~}\jref{\bf #1}{#2}{#3}}}

\def\zpc#1#2#3#4{\href{https://dx.doi.org/#4}{{\it Z.~Phys.~}\jref{\bf C #1}{#2}{#3}}}

\def\otherjournal#1#2#3#4#5{\ifthenelse{\equal{#5}{}}{{\it #1}\jref{\bf #2}{#3}{#4}}{\href{https://dx.doi.org/#5}{{\it #1}\jref{\bf #2}{#3}{#4}}}}
\def\proc#1#2{\ifthenelse{\equal{#2}{}}{in proceedings of {\it #1}}{in proceedings of \href{https://dx.doi.org/#2}{\it #1}}}
\newcommand{\jref}[3]{{\bf #1} (#2) #3}
\newcommand{\hepph}[1]{\href{https://arXiv.org/abs/hep-ph/#1}{\texttt{arXiv:hep-ph/#1}}}

\newcommand{\arxiv}[2]{\href{https://arXiv.org/abs/#1}{\texttt{arXiv:#1\,[#2]}}}
\newcommand{\bibentry}[4]{#1, {\it #2}, #3\ifthenelse{\equal{#4}{}}{}{, }#4.}

\title{\texttt{SusHi 2.0}\\Higgs production cross sections in BSM models}
\ShortTitle{\texttt{SusHi 2.0}}

\author*[a]{S.Y.~Klein}

\affiliation[a]{Institute for Theoretical Particle Physics and
  Cosmology,\\ RWTH Aachen University, Sommerfeldstraße 16, 52074 Aachen, Germany}

\emailAdd{yklein@physik.rwth-aachen.de}

\abstract{
  A new upcoming version of \sushi\ is introduced. It features unified input
  for the \sm\ and \bsms\ parameters for higher-order total cross
  sections for Higgs production in gluon fusion, heavy-quark
  annhilation, as well as Higgsstrahlung. Like previous versions of
  \sushi, it provides links to codes like \thdmc\ and \feynhiggs, but
  can also process standard SLHA output of spectrum generators like
  \softsusy\ and \spheno.
}

\FullConference{The European Physical Society Conference on High Energy Physics (EPS-HEP2023)\\
 21-25 August 2023\\
Hamburg, Germany\\}

\renewcommand{\logo}{\relax}

\reportnumber{TTK-23-30 / P3H-23-085}

\begin{document}

\input{./sections/content.tex}

\end{document}

%% file: sections/content.tex
\maketitle

\section{Introduction}

After the discovery of the Higgs boson in 2012 at the \lhc\ by the ATLAS and
CMS collaborations \cite{ATLAS:2012yve,CMS:2012qbq}, the Higgs boson remains
to be an interesting probe for precision tests of the \sm\ and the search
for \bsms. Therefore, in addition to precise measurements,
precise predictions for Higgs observables in the \sm\ and also beyond are
required. Important processes to consider are single Higgs production and the
production of a Higgs boson in association with a vector boson, so called
Higgsstrahlung. Here, we want to give an update on the development of an
upcoming version of \sushi\
\cite{Harlander:2012pb,Harlander:2016hcx,sushi}, which will include
cross-section predictions for single Higgs production in gluon fusion
and heavy-quark annihilation and for Higgsstrahlung, the inclusion of
the latter being a new feature realised by merging \vh\
\cite{Brein:2012ne,Harlander:2018yio,vhnnlo} into \sushi.


\section{Higgs production in gluon fusion}

The main production mode for the Higgs boson in the \sm\ is gluon
fusion. In \sushi\ it is implemented via the subprocedure \ggh. At \lo\ only
two diagrams contribute to the process, where the Higgs boson is coupled to
the two gluons via a heavy-quark triangle loop, see \fig{gghdia}. The partonic cross
section at this order in the \sm\ has been obtained in
\citere{Georgi:1977gs}
At \nlo, the partonic cross sections are still expressible analytically in the
full \sm\ \cite{Djouadi:1991tka,Harlander:2005rq} and are implemented
in \sushi\ with exact quark mass dependence. At \nnlo\ and \ntlo, partonic
cross sections are implemented in the \htl, meaning that the \lo\ cross
section is rescaled by the corresponding perturbative K-factor which is
evaluated in the limit of an infinitely heavy top
quark~\cite{Harlander:2002wh,Anastasiou:2002yz,Ravindran:2003um,Anastasiou:2011pi,Anastasiou:2015vya}. At 
\nnlo\ (and \nlo), comparison
to the exact calculation\,\cite{Czakon:2021yub} shows that this provides an
excellent approximation at the few-permil level.


The cross section can also be obtained in various \bsms, for example in
the \thdm. In the \thdm\ \sushi\ can not only compute cross section
predictions for the production of the \sm-like Higgs but also for the
other uncharged Higgs bosons. This is achieved by adjusting the Yukawa
coupling of the massive quarks to the Higgs. In the \mssm, also squark
and gluino contributions are taken into account. \sushi\ uses results
obtained in \citeres{Spira:1995rr,Ravindran:2003um, Harlander:2003bb,
Harlander:2004tp,Degrassi:2008zj,Harlander:2003kf, Degrassi:2011vq,
Degrassi:2012vt} for the partonic cross sections in the \mssm.



\section{Higgs production in heavy-quark annihilation}

\begin{figure}
\phantom{|}\hfill
\begin{subfigure}[c][][c]{0.25\textwidth}
\def\svgwidth{\textwidth}\input{./diagrams/ggh-lo.tex}
\caption{\lo\ diagram for Higgs production in gluon fusion.\\\phantom{.}}
\label{gghdia}
\end{subfigure}\hfill
\begin{subfigure}[c][][c]{0.25\textwidth}
\def\svgwidth{\textwidth}\input{./diagrams/bbh-lo.tex}
\caption{\lo\ diagram for Higgs production in heavy-quark annihilation.}
\label{bbhdia}
\end{subfigure}\hfill
\begin{subfigure}[c][][c]{0.25\textwidth}
\def\svgwidth{\textwidth}\input{./diagrams/vh-lo.tex}
\caption{\lo\ diagram for Higgsstrahlung.\\\phantom{.}}
\label{vhdia}
\end{subfigure}\hfill
\phantom{|}
\caption{\lo\ diagrams for the processes available in \sushit\ drawn with FeynGame \cite{Harlander:2020cyh}.}
\label{diagrams}
\end{figure}
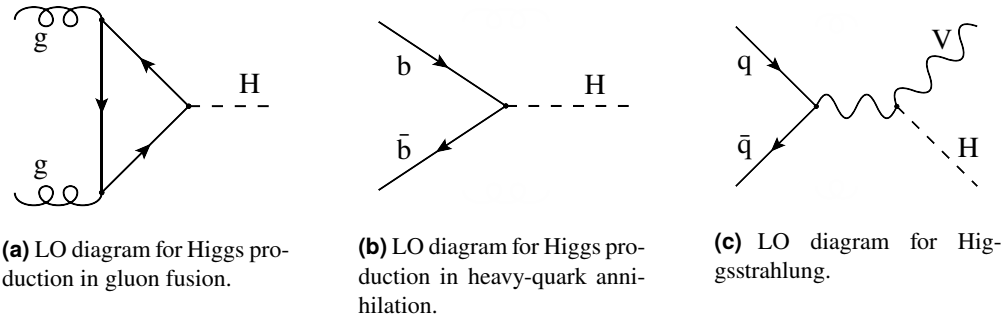

\bsms\ can enhance the coupling of the Higgs-like particles to the
bottom quark, while possibly decreasing its coupling to the top quark
at the same time. Therefore the production of a Higgs in bottom-quark
annihilation can become sizeable. \sushi\ includes this process via
the subprocedure \bbh. Also included is Higgs production in the
annihilation of any two initial state quarks as described in
\citere{Harlander:2015xur}. The cross-section predictions for these
processes can also be obtained in the \mssm\ and \thdm. Here the \sm\
result is re-weighted by the Yukawa coupling of the heavy quark to the
Higgs-like particle. An example Feynman diagram is shown in \fig{bbhdia}.

\section{Higgsstrahlung}

The calculation of Higgsstrahlung cross-section predictions will be
included in the upcoming version of \sushi. These predictions are part
of \vh\ which will be merged into \sushi. Higgsstrahlung is the
production of a vector boson in association with a Higgs boson:
$\text{pp}\rightarrow \text{VH},\,\text{V}\in\lbrace W^+,W^-,Z\rbrace$.
Up to \nlo\ this process appears only in a \dy-like fashion: A virtual
vector boson $\text{V}^*$ is produced which subsequently radiates off a
Higgs boson (hence the name "Higgsstrahlung"), as shown in \fig{vhdia}. These contributions can
be calculated by integrating the product of the hadronic cross section
for the production of the intermediate virtual vector boson
$\sigma_{\text{pp}\rightarrow\text{V}^*}$ with the decay width of this
intermediate particle to the final state $\text{d}
\Gamma_{\text{V}^{*}\rightarrow\text{VH}}/\text{d}q^2$:
\begin{equation}
\label{eq:dy}
\sigma_{\text{pp}\rightarrow\text{VH}} = \int_{\left(m_\text{H}+
m_\text{V}\right)^2}^s \text{d}q^2 \; \sigma_{\text{pp}
\rightarrow\text{V}^{*}} \; \frac{\text{d}\Gamma_{\text{V}^{*}
\rightarrow\text{VH}}}{\text{d}q^2}\text{.}
\end{equation}
\sushi\ builds on \zwprod\ \cite{zwprod} for the calculation of the hadronic
cross section of the production of the intermediate state which uses results
obtained in \citeres{Hamberg:1990np,Harlander:2002wh}.

At \nnlo, also other contributions have to be taken into account. The Higgs
can couple to a top loop instead of an intermediate vector boson.  In
addition, for ZH production, also the gluon fusion channel,
$\text{gg}\rightarrow\text{ZH}$, has to be taken into account.

\bsm\ effects are included in a similar fashion to the Higgs production
case in \sushi. In the \thdm, production of all uncharged Higgs bosons
together with the vector boson is allowed by adjusting the couplings of
the Higgs to the other particles. In the \mssm\ squarks have to be
considered for the gluon-initiated contribution. Additionally, in both
the \thdm\ and the \mssm, a Higgs boson can be produced which then
decays into a vector boson and a Higgs with opposite parity.


\section{Usage of \sushi}

The compilation of \sushi\ requires \lhapdf\ \cite{Whalley:2005nh,
Bourilkov:2006cj,Buckley:2014ana,lhapdf} to be linked against it for the
availability of \pdf\ sets. This in turn means that all \lhapdf\ compatible
\pdf\ sets can be used in \sushi. The new
version will also require linkage to the \cuba\
\cite{Hahn:2004fe,Hahn:2014fua,cuba} library for \vegas\
\cite{Lepage:1978} \moca\ integration and the \lt\ library
\cite{Hahn:1998yk,Oldenborgh:1990,looptools} for the evaluation of box
integrals in the gluon-gluon initiated contribution to ZH production.

Building \sushi\ results in an executable file which has to be called with a
command line argument pointing to an input file. This input file follows
the \slha\ \cite{Skands:2003cj,Allanach:2008qq} format.  If specified, \sushi\
will use a spectrum generator to generate input parameters not found in the
input file. Afterwards, \sushi\ computes the cross sections requested in the
input file and writes the results into an output file which also follows
the \slha\ format.

In the remainder of this section, we first describe the input file of
\sushi\ followed by an overview of the usage of spectrum generators.
Finally, the output file is discussed.


\subsection{Input}

The instructions for the calculations that the user wants \sushi\ to perform
have to be written into the input file. This includes specifying what
processes up to which orders are to be calculated. In the Higgsstrahlung case,
also the different contributions can be turned on and off. One has to specify
which collider, $\text{pp}$ or $\text{p}
\bar{\text{p}}$, at which centre of mass energy as well as the PDF set
to be used amongst other things. The input has to follow the \slha\
format as shown in \fig{code:slha}. Essentially, parameters are grouped
together in named \emph{blocks}. These blocks are initiated by a line
starting with the keyword \textbf{\texttt{Block}} followed by the name
of the block. In the subsequent lines, the parameters belonging to this
block are given by first stating the index, or in some cases indices, of
the parameter and then the input value for this parameter. On the left
side of \fig{code:slha}, index~\texttt{1} of the block \texttt{PDFSPEC}
denotes the \pdf\ set to be used via \lhapdf\ while index \texttt{10}
indicates the \pdf\ set member.

\begin{figure}[H]
\begin{minipage}{0.5\textwidth}
\label{code:slha_input}
\lstset{
  emph={Block},emphstyle=\textbf,
  emph={[2]d},emphstyle={[2]\underline},
  morecomment=[l]{\#}
}
\begin{lstlisting}
Block VEGAS
  10   10000
  11    5000
Block PDFSPEC
   1 PDF4LHC21_40_pdfas
  10      0
\end{lstlisting}
\end{minipage}%
\begin{minipage}{0.5\textwidth}
\label{code:slha_output}
\lstset{
  emph={Block},emphstyle=\textbf,
  emph={[2]d},emphstyle={[2]\underline},
  morecomment=[l]{\#}
}
\begin{lstlisting}
Block SIGMA
   1 5.32615250E-01
 111 2.02093466E-04
  10 1.18000798E-01
  11 5.32615250E-01
 110 2.02093466E-04
\end{lstlisting}
\end{minipage}
\caption{Excerpts of sample input (left) and output (right)
files in the \slha\ format.}
\label{code:slha}
\end{figure}

In the upcoming version of \sushi, indices and/or blocks of some parameters
will be changed compared to the current version in order to avoid conflicts
with the input scheme for \vh.


\subsection{Spectrum generators}

Spectrum generators can be used to calculate Higgs masses and couplings,
squark masses and other parameters used by \sushi\ in various \bsms\
from input containing more fundamental parameters of the theory. In
particular \feynhiggs\ \cite{Heinemeyer:1998np,Heinemeyer:1998yj,
Degrassi:2002fi,Frank:2006yh,Hahn:2013ria,Bahl:2016brp,Bahl:2017aev,
Bahl:2018qog,feynhiggs} and \thdmc\ \cite{Eriksson:2009ws,thdmc} are
tightly integrated and can be used straightforwardly in \sushi\ after
linking against them.

\feynhiggs\ and \thdmc\ are activated by including the blocks
\texttt{FEYNHIGGS} and \texttt{2HDMC} respectively. These blocks have to
contain all the relevant input parameters for the given spectrum
generator. There are two ways that \sushi\ interacts with these two spectrum
generators. The library functions of the packages can either be used directly,
or \sushi\ uses the command line modes.  In the latter case, for \feynhiggs,
an input file is created by \sushi\ which is subsequently read by \feynhiggs,
whose output file is then read by \sushi. For \thdmc, this program is called
with the parameters specified directly in the command
line. Nevertheless, \thdmc\ also generates an output file in the \slha\ format
which is subsequently read by \sushi.

Support for \fs\ \cite{Athron:2014yba,Athron:2017fvs,flexiblesusy}
and \texttt{Himalaya} \cite{Harlander:2017kuc} is also directly built
into \sushi, albeit only in a command line mode.  If the input file contains
the block \texttt{FLEXIBLESUSY}, the whole input file gets converted into
a \fs-compliant file which is then used as an input
for \fs. The \slha-formatted output is then read by
\sushi, and the parameters that were calculated by \fs\ are used in the
cross section prediction calculations.

Other spectrum generators like \softsusy\ \cite{Allanach:2001kg,
softsusy} and \spheno\ \cite{Porod:2003um,Porod:2011nf,spheno} can be
used as well. These generators have to be called manually such that they
produce an output file in the \slha\ format. This file can then be read
by \sushi\ by specifying the filename in the \texttt{SPECTRUMFILE}
block in its input. One can of course also create such a spectrum file
by hand and use it in the same way.


\subsection{Output}

After processing the input and possibly calling spectrum generators to
gather all the relevant parameters, \sushi\ calculates the requested
cross-section predictions. The results are written into an output file
following the \slha\ format, just like the input file. This means that
the results each belong to a specific index in a specific block. In
\fig{code:slha}, for example, index \texttt{11} of block \texttt{SIGMA}
denotes the \dy-like contribution to the Higgsstrahlung cross section at
\nnlo, and index \texttt{110} its \moca\ integration error.

The output file also contains a reconstruction of the input file after
the results of the calculation(s). Therefore, the output file can be
used as an input file. This is useful for testing reproducibility but
also has the practical advantage of not having to keep track of which
output file belongs to which input parameter set.




\section{Drell-Yan-like Higgsstrahlung at \ntlo}

The \dy\ part is the most important contribution to the Higgsstrahlung cross
section as it is the only contribution that appears at
\lo\ and \nlo.
As shown in \eqn{eq:dy}, two components are needed for the calculation of this
contribution: the hadronic cross section for the production of the virtual
vector boson and its decay width to
the VH final state. The decay
width does not entail pure \qcd\ corrections and therefore only the
cross section for the production of the intermediate state has to be adjusted when going
to higher orders. The hadronic cross section in turn is obtained from the
partonic cross section via convolution with the \pdfs.
Therefore, corrections on the partonic cross sections for
the production of an intermediate vector boson are necessary for higher orders
in the \dy\ contribution to Higgsstrahlung. These are available up
to \nnlo\ \cite{Brein:2003wg} and \ntlo\
\cite{Duhr:2020sdp,Duhr:2021vwj}. Their implementation in \sushi\
is a new feature of the upcoming version of \sushi.

For the \dy\ part of the cross section it is straightforward to
calculate not only total cross sections but also distributions in the
invariant mass of the VH system. There are two possibilities for the
calculation starting from \eqn{eq:dy}: On the one hand, one can simply
omit the integration over the virtuality of the intermediate vector
boson. This yields the differential distribution:
\begin{equation}
\label{eq:diff_distrib}
\frac{\mathrm{d}\sigma}{\mathrm{d}m_\text{VH}^2} = \sigma_{\text{pp}
\rightarrow\text{V}^{*}}\; \frac{\text{d}\Gamma_{\text{V}^{*}
\rightarrow\text{VH}}}{\text{d}m_\text{VH}^2}\text{.}
\end{equation}
This is the method used in \vh. On the other hand, one can, instead of
getting rid of the integration altogether, simply restrict the range of
the integration over the virtuality to bins. One gets a binned
distribution:
\begin{equation}
\label{eq:bin_distrib}
\delta \sigma_{\text{pp}\rightarrow\text{VH}} =
\int_{m_\text{VH}^2-\frac{1}{2}\delta
m_\text{VH}^2}^{m_\text{VH}^2+\frac{1}{2}\delta m_\text{VH}^2}
\text{d}q^2 \; \sigma_{\text{pp}\rightarrow\text{V}^{*}}\;
\frac{\text{d}\Gamma_{\text{V}^{*}\rightarrow\text{VH}}}
{\text{d}q^2}\,,
\end{equation}
where $\delta m_\text{VH}^2$ denotes the bin width. The binned
distribution is added to the new version of \sushi\ although also the
differential distribution will be available at up to \ntlo.

\begin{figure}[H]
\begin{center}
\includegraphics[width=0.8\textwidth]{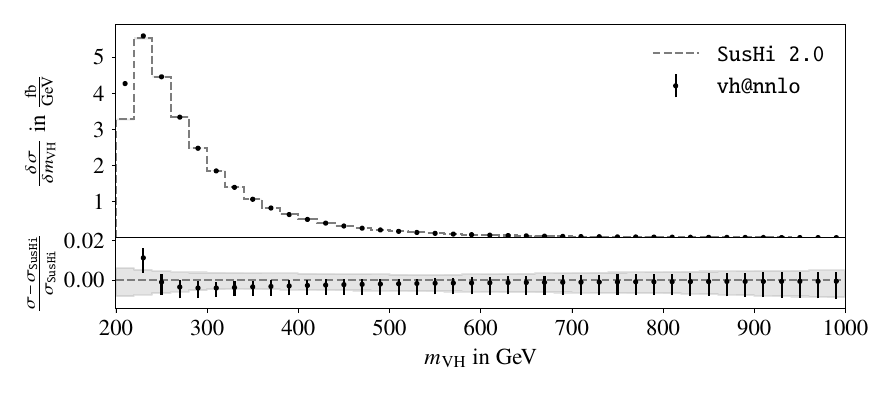}
\end{center}
\caption{Distribution of the cross section for the \dy-part
of the Higgsstrahlung $\text{pp}\rightarrow\text{W}^-\text{H}$ cross
section in the invariant mass of the W$^-$H system at \nnlo. The dots with the error bars indicate the values
obtained with \vh. The dashed line and the error band is the binned distribution calculated
with the new implementation. The errors are given by a 7 point scale
variation with a factor of 3.}
\label{plot:dy_dist}
\end{figure}

A comparison between the two approaches can be seen in
\fig{plot:dy_dist}. The differences arise because
the binned distribution consists of the averages of the differential
distribution over the bins, while \vh\ evaluates the differential distribution
according to \eqn{eq:diff_distrib} at the centre of each bin. The new
implementation gives the averages according to
\eqn{eq:bin_distrib}, which are expected to differ from the values at the
centres of the bins.








\section{Summary}

The upcoming version of \sushi\ will include Higgsstrahlung
cross-section prediction in addition to single Higgs production as it
incorporates \vh. Therefore \vh\ and \sushi\ will be merged into a
single package. For the Higgsstrahlung process \ntlo\ corrections for
the \dy-part will be taken into account.

\section*{Acknowledgements}

This research was supported by the Deutsche Forschungsgemeinschaft (DFG,
German Research Foundation) under grant 396021762 - TRR 257, and by BMBF
under grant 05H21PACCA

%% file: diagrams/ggh-lo.tex
\begingroup%
  \makeatletter%
  \providecommand\rotatebox[2]{#2}%
  \ifx\svgwidth\undefined%
    \setlength{\unitlength}{595.32000732bp}%
    \ifx\svgscale\undefined%
      \relax%
    \else%
      \setlength{\unitlength}{\unitlength * \real{\svgscale}}%
    \fi%
  \else%
    \setlength{\unitlength}{\svgwidth}%
  \fi%
  \global\let\svgwidth\undefined%
  \global\let\svgscale\undefined%
  \makeatother%
  \parbox[c][][c]{\unitlength}{%
    \begin{picture}(1,0.8047808764940239)%
      \put(0,0){\includegraphics[width=\unitlength,viewport = 0 0 450 362]{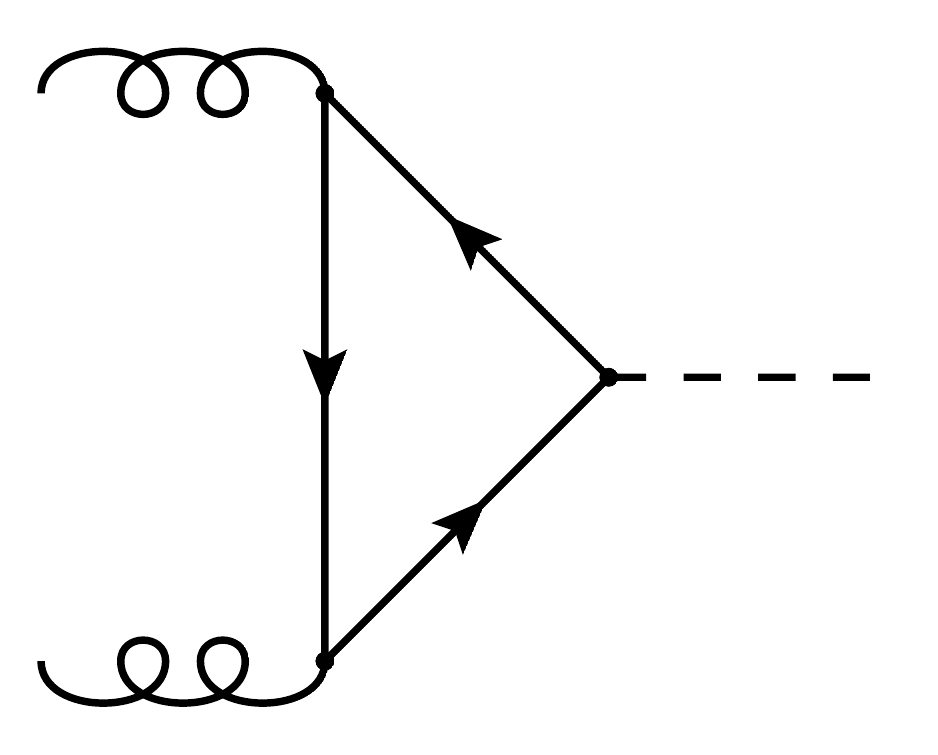}}%
      \put(0.13466135819119762,0.6254980079681275){\raisebox{-0.5\height}{\rotatebox{0.0}{\scalebox{1.0}{\makebox(0,0)[]{{$\text{g}$}}}}}}%
      \put(0.13466135819119762,0.17928286852589642){\raisebox{-0.5\height}{\rotatebox{0.0}{\scalebox{1.0}{\makebox(0,0)[]{{$\text{g}$}}}}}}%
      \put(0.8613545780637825,0.4820717131474104){\raisebox{-0.5\height}{\rotatebox{0.0}{\scalebox{1.0}{\makebox(0,0)[]{{$\text{H}$}}}}}}%
    \end{picture}%
  }%
\endgroup%

%% file: diagrams/bbh-lo.tex
\begingroup%
  \makeatletter%
  \providecommand\rotatebox[2]{#2}%
  \ifx\svgwidth\undefined%
    \setlength{\unitlength}{595.32000732bp}%
    \ifx\svgscale\undefined%
      \relax%
    \else%
      \setlength{\unitlength}{\unitlength * \real{\svgscale}}%
    \fi%
  \else%
    \setlength{\unitlength}{\svgwidth}%
  \fi%
  \global\let\svgwidth\undefined%
  \global\let\svgscale\undefined%
  \makeatother%
  \parbox[c][][c]{\unitlength}{%
    \begin{picture}(1,0.7859922178988327)%
      \put(0,0){\includegraphics[width=\unitlength,viewport = 0 0 450 354]{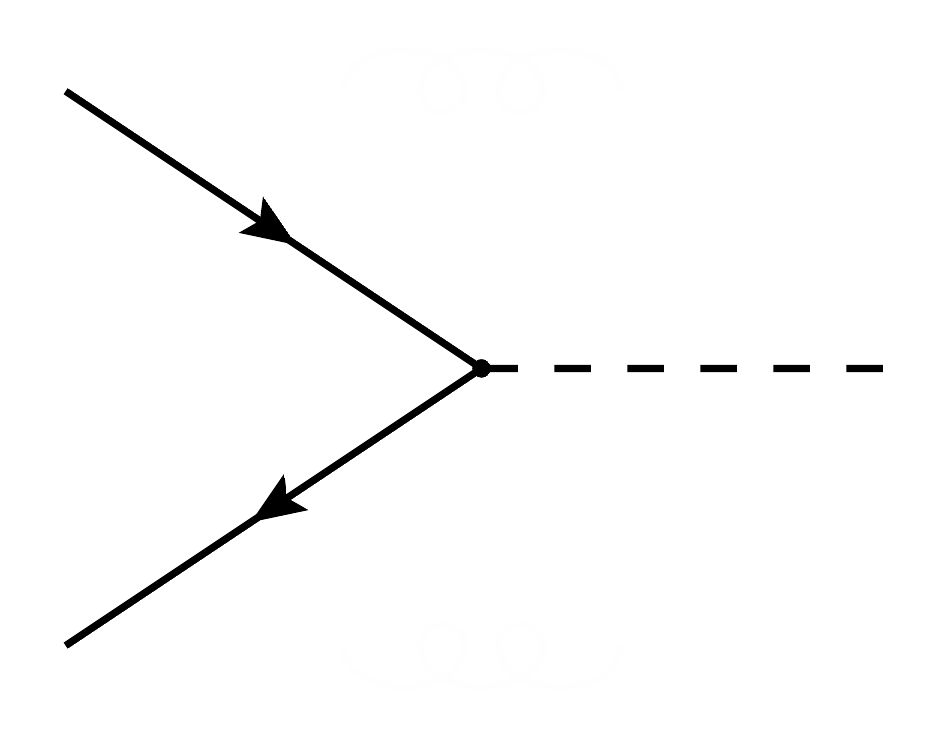}}%
      \put(0.8241245083307942,0.4708171206225681){\raisebox{-0.5\height}{\rotatebox{0.0}{\scalebox{1.0}{\makebox(0,0)[]{{$\text{H}$}}}}}}%
      \put(0.15994551530926585,0.53524900080643){\raisebox{-0.5\height}{\rotatebox{0.0}{\scalebox{1.0}{\makebox(0,0)[]{{$\text{b}$}}}}}}%
      \put(0.15994551530926585,0.25074321709240266){\raisebox{-0.5\height}{\rotatebox{0.0}{\scalebox{1.0}{\makebox(0,0)[]{{$\bar{\text{b}}$}}}}}}%
    \end{picture}%
  }%
\endgroup%

%% file: diagrams/vh-lo.tex
\begingroup%
  \makeatletter%
  \providecommand\rotatebox[2]{#2}%
  \ifx\svgwidth\undefined%
    \setlength{\unitlength}{595.32000732bp}%
    \ifx\svgscale\undefined%
      \relax%
    \else%
      \setlength{\unitlength}{\unitlength * \real{\svgscale}}%
    \fi%
  \else%
    \setlength{\unitlength}{\svgwidth}%
  \fi%
  \global\let\svgwidth\undefined%
  \global\let\svgscale\undefined%
  \makeatother%
  \parbox[c][][c]{\unitlength}{%
    \begin{picture}(1,0.7453874538745388)%
      \put(0,0){\includegraphics[width=\unitlength,viewport = 0 0 450 335]{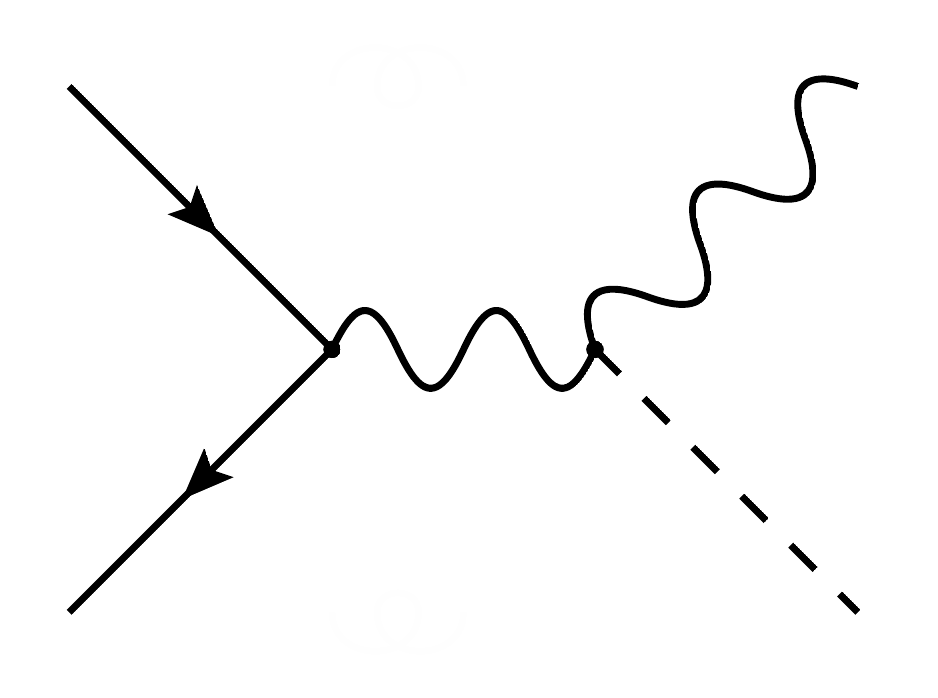}}%
      \put(0.7918575548937682,0.6081424384199543){\raisebox{-0.5\height}{\rotatebox{0.0}{\scalebox{1.0}{\makebox(0,0)[]{{$\text{V}$}}}}}}%
      \put(0.8831813089215511,0.22856876948236735){\raisebox{-0.5\height}{\rotatebox{0.0}{\scalebox{1.0}{\makebox(0,0)[]{{$\text{H}$}}}}}}%
      \put(0.1057485803773419,0.22856876948236735){\raisebox{-0.5\height}{\rotatebox{0.0}{\scalebox{1.0}{\makebox(0,0)[]{{$\bar{\text{q}}$}}}}}}%
      \put(0.10574858037734187,0.5168186843921714){\raisebox{-0.5\height}{\rotatebox{0.0}{\scalebox{1.0}{\makebox(0,0)[]{{$\text{q}$}}}}}}%
    \end{picture}%
  }%
\endgroup%